# Concerns and Limitations in Agile Software Development: A Survey with Paraguayan Companies

Myrian R. N. Salinas[1], Adolfo Gustavo Serra Seca Neto[1], Maria Claudia F. P. Emer[1]

[1] Postgraduate Program in Applied Computing
Academic Department of Informatics
Federal University of Technology – Parana – Brasil
`michinoguera@gmail.com, adolfo@utfpr.edu.pr, mclaudia@-dainf.ct.utfpr.edu.br`

**Abstract.** This year, the Agile Manifesto completes seventeen years and, throughout the world, companies and researchers seek to understand their adoption stage, as well as the benefits, barriers, and limitations of agile methods. Although we have some studies and questionnaire data at the global level, we know little about how the Paraguayan software community is adopting agile methods. The present work conducted a research to set up the current stage of adoption, initial concerns and barriers of implementation of agile methods in software development companies in Paraguay. An online survey was sent to representatives of 53 Paraguayan companies. Of these, 9 (17%) companies responded. The concern about adopting more agile methods (44.44% of respondents) was the lack of reliability in product quality if developed using agile methods. The main barrier was the lack of experience (66.66%) of the companies.

## 1 Introduction

Agile Software Development (ASD) was formally presented to the software engineering community in 2001 through a document called "Agile Manifesto", which mentions a set of core values and principles that emphasized Agility, in other words, The ability to adapt to fast volatile requirements [1]. However, agile principles don't suggest specific activities or artifacts; these are defined in a number of methods and practices such as Scrum, Extreme Programming (XP), Test Driven Development, Lean Software Development, Kanban etc. Practices vary and focus on different aspects of agile principles and address different problems in software development.

From this, development with agile methods has attracted the attention of many researchers. Most of the available studies report experiences, generally positive, with their application in specific organizations and projects and, therefore, hardly generalizable.

Considering their current popularity of agile methods, the interest of the first author of Paraguayan nationality and available literature on studies done elsewhere, it's relevant to investigate the adoption, barriers, and limitations with respect to agile methods. This work has as scope the domains mentioned above in the Paraguayan context of software development with the help of a research. The remainder is organized as follows. Section 2 is the literature review followed by Section 3, which outlines objectives and research methodology. In Section 4, we analyse the results, and Section 5 presents the conclusion..

## 2    Literature Review

The term "Agile Methodologies" emerged in 2001, when a group of software development process specialists decided to meet in the US to discuss ways to improve the performance of their projects, and wrote a document called *The Agile Manifesto*. Methods and practices TDD [21], Pair Programming [22] e Planning Poker [23] related to this manifesto have been increasingly adopted in recent years.

Several authors have pointed out the advantages of agile methods, with their emphasis on individuals and process interactions, client collaboration on formal contracts and negotiations, and responsiveness to rigid planning [8, 9, 10, 11, 12, 15, 16, 17, 18]. However, there are few studies on adoption difficulties [8, 13, 14, 19, 20].

A survey conducted by VersionOne in 2016 suggested the main difficulties in adopting agile methods: organizational culture in disagreement with agile values, (63%) and lack of skills or experience with agile methods (47%).

Another research [3, 13] in agile methods was conducted in 2013 by a group from the University of São Paulo, to set up the current stage of adoption and adaptation of agile methods in Brazil. The results showed that the main concern in adopting the method was the lack of documentation. In addition, the major barrier to broad adoption was the ability to change organizational culture.

In February 2015, both Gartner and Software Advice [4, 5] launched research and analysis on agile life-cycle management or project management tools. Of the project managers who responded, 49% say that coaching others is a common challenge they face, especially adopting agile culture.

Another literature review study [6] focused on the current challenges of this agile stream. The most significant were team management, agility in distributed teams, prioritization of requirements, documentation, change requirement, organizational culture, process and monitoring, and feedback.

## 3 Objetives and Methodology

### 3.1 Definition of goals

The main objective of our study was to set up the current adoption stage, barriers and limitations regarding the use of agile methods in software development companies in Paraguay.

### 3.2 Methodology

For the accomplishment of the study a research was prepared by means of an online survey. The following are the steps performed in the study (Fig. 7):

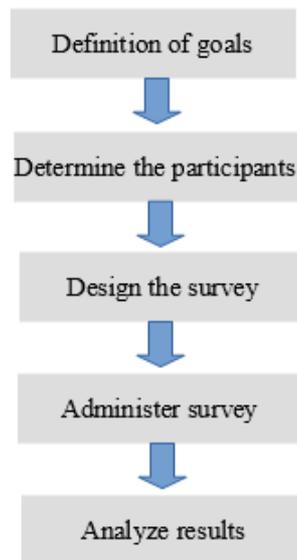

**Fig. 7.** Methodology adopted (Own authorship)

**Determine the participants.** A common problem when conducting an online survey is finding the right respondents and collecting enough answers so that you have relevant data. Our primary concern, therefore, was to find the right respondents, whose response is valuable enough to analyse the end result as managers and development managers. In our research, the questionnaire was disseminated directly to the directors or development managers of the companies.

According to the list provided by the Directorio de la Red de Inversiones y Exportaciones (REDIEX), which belongs to the Ministerio de Industria y Comercio de Paraguay, there are 53 companies registered in the Software Development category. The questionnaire was sent to all the companies on the list and 9 of them answered.

**Design the survey.** We adopted the creation of an online questionnaire that consisted of ten multiple-choice questions.

The first section of this survey has general information. The details sought include the name of the organization to which the respondent belongs, the position, how many people in total are employed in the company.

The second section deals with the adoption of agile methods, in which the questions were structured in such a way as to answer the main issues of adoption: concerns and barriers. The questions were, for instance, how many years of experience do you have using agile methods (to understand the extent of company familiarity with agile development) and what were the difficulties of adoption (to identify the reasons).

The last section complements with the percentage information of projects developed with agile methods.

**Administer survey.** The research survey was directly disclosed to the directors or development managers of the companies through an e-mail, to which they responded by filling out the online questionnaire.

The participants were mainly representatives who had full knowledge of the company policies, the various methods used and the time the company has using process development.

**Analyze results.** The analysis of the results was based on the answers that we received through the online questionnaire. Responses were carefully analysed in order to get accurate results based on the research. The main concern was to interpret the information in the wrong way, which would definitely not serve our purpose of investigation.

## 4	Results

The data collected with the help of the form gave us a clear idea of the respondent and his position. Most of the participants are Project Managers and President of the Company, 33.33% in both cases, which ensures a responsible and official response (Fig. 8) and also confirms the current use of agile methods of 100% of the participants (Fig. 9).

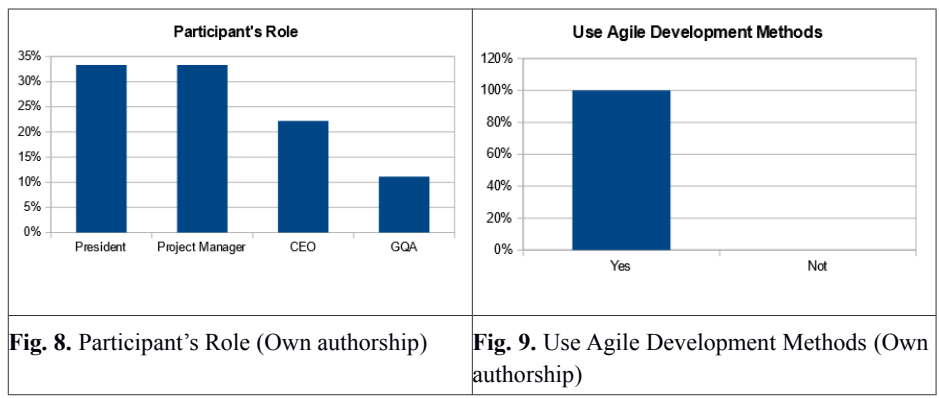

**Fig. 8.** Participant's Role (Own authorship)    **Fig. 9.** Use Agile Development Methods (Own authorship)

Another important feature is the size of the software development team. Most (66.66%) of the companies have up to 20 employees in their team (Fig. 10).

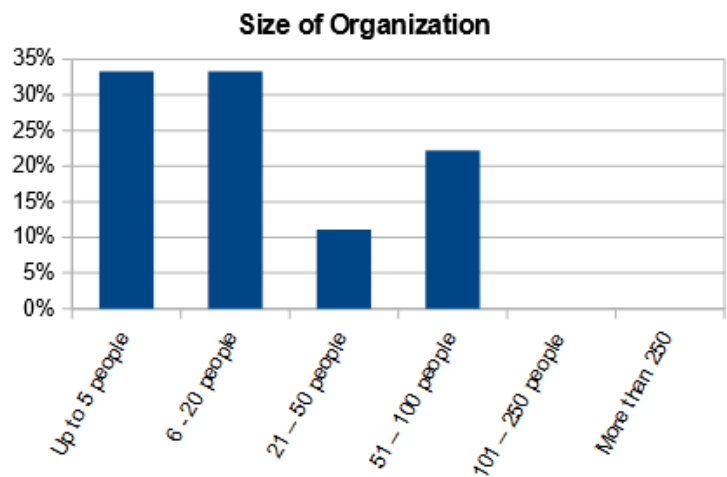

**Fig. 10.** Size of Organization (Own authorship)

One of the main themes of this research details the following concerns (Fig. 11):

- Inability to scale: Corresponds to the lack of organizational capacity to make the shift to agile method.
- Reduced software quality: It's the perception or lack of reliability in delivering a quality product or ensuring customer satisfaction.
- Dev team opposed to change: Occurs when developers are not convinced or motivated to make the move to agile methods.
- Lack of early planning: When participants are unaware of the activities needed to make the change because of lack of planning.
- Internal company regulations: When the rules or rules of the company don't conform to the principles of the methods.
- No concerns: They had no concerns about adopting the methods.

The data show that 44.44% of the participants had concerns about software quality when adopting the new method. Other significant reasons are: inability to escalate, with 22.22% and development team resistant to changes, with 22.22%.

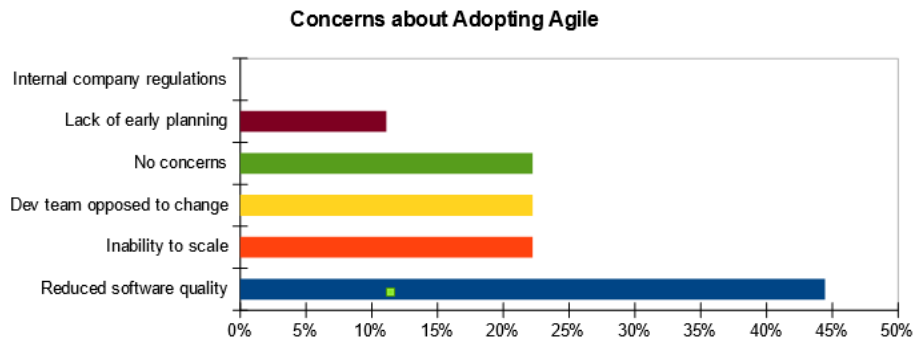

**Fig. 11.** Concerns about Adopting Agile (Own authorship)

Other important themes in the research are identifying the barriers to futher adoption in the enterprise (Fig. 12). The reasons are detailed as follows:

- Company's internal rules or standards: When the company's rules don't match with the principles of the method.
- Budget constraints: The company has no budget for the broad adoption, but it has already implemented agile methods in some of its projects.

- Project complexity: The company also works with large and complex projects and uses agile methods to develop small projects.
- Customer collaboration: The client has no interest in participating in meetings and other activities appropriate to the agile methods or techniques used.
- Confidence in the ability to scale: Corresponds to difficulties to make the change to agile method in order to increase its scale. That is, the difficulty in using agile methods in more projects and/or bigger projects.
- Lack experience: The team does not have sufficient experience for the wide adoption of agile methods.
- Other: Other reasons not mentioned on the list.
- None: They had no barriers in adopting agile methods.

The factors which are main chosen as main barriers to adoption of the total agile (Fig. 12) are: a) lack experience with 66.66%, b) project complexity, 33.33% c) customer collaboration, with 33.33% and d) confidence in the ability to scale with 33.33%.

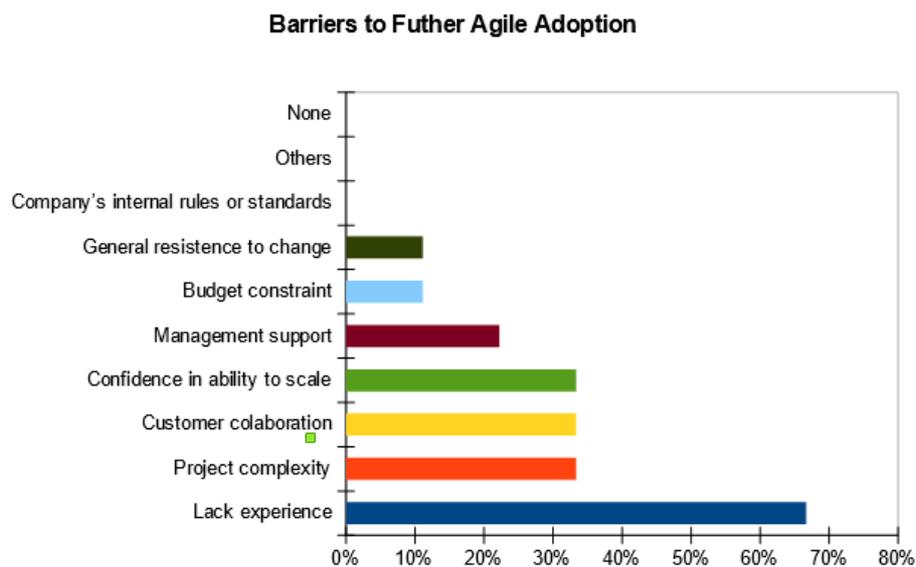

**Fig. 12.** Barriers to futher Agile Adoption (Own authorship)

Experience time is an important factor for the wide adoption of agile methods. The majority (55.56%) of the participating companies have average experience of 1 to 2 years (Fig. 13).

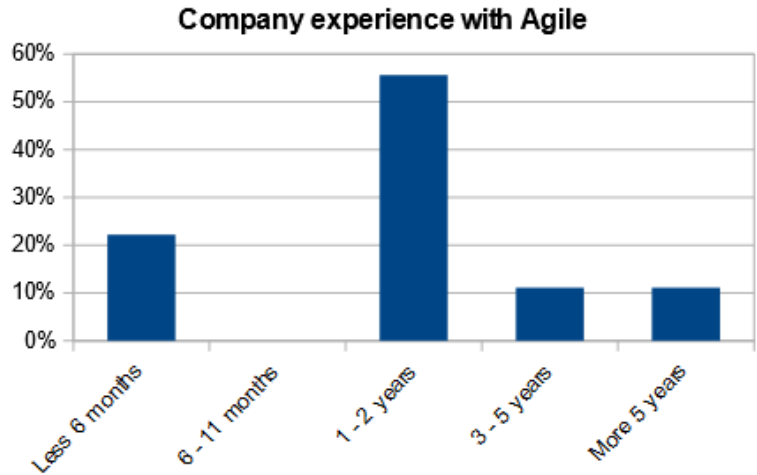

**Fig. 13.** Company experience with Agile (Own authorship)

The choices of methods and techniques are also fundamental according to the knowledge, the characteristics of the team and the company (Fig. 14 e Fig. 15). Most of the companies interviewed prefer Scrum and the most used practices are: Unit tests with 55.56%, Short iterations with 44.44%, Backlogs prioritized with 33.33%, Daily meeting with 22, 22%, Retrospectives with 22,22%, Release planning with 22,22%, Continuous integration with 22,22% and Open work area with 22,22%.

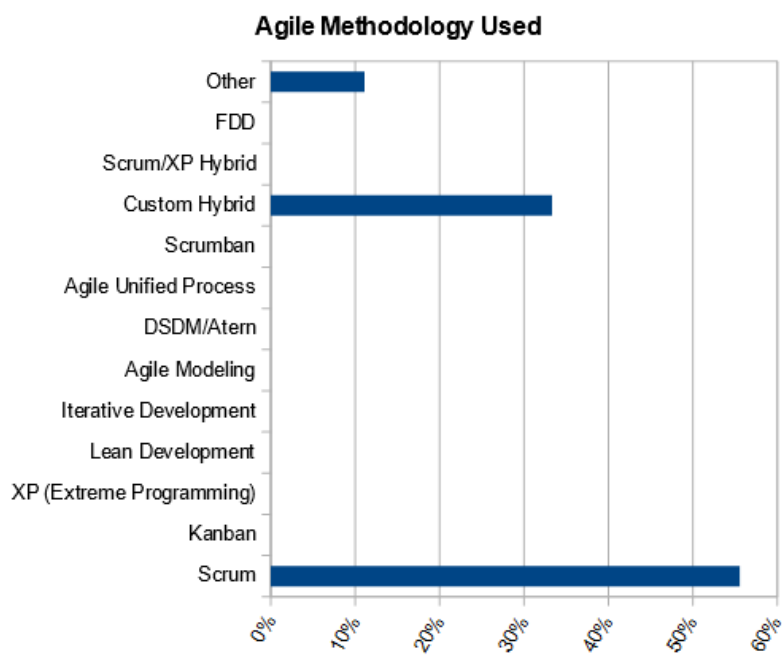

**Fig. 14.** Agile Methodology used (Own authorship)

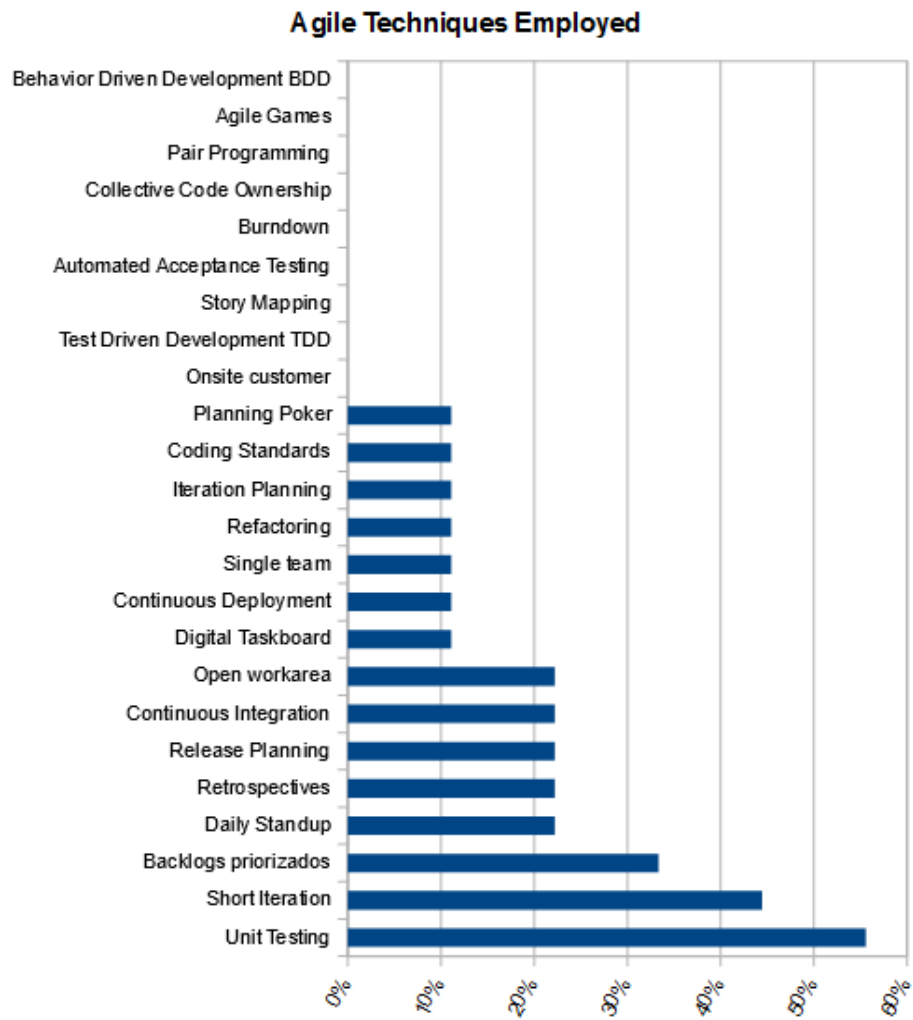

**Fig. 15.** Agile Techniques Employed (Own authorship)

Another data that allows us to visualize the adoption level is the quantity of projects developed with agile methods (Fig. 16). The majority (55.56%) of the companies used agile methods in 50% or more of their projects.

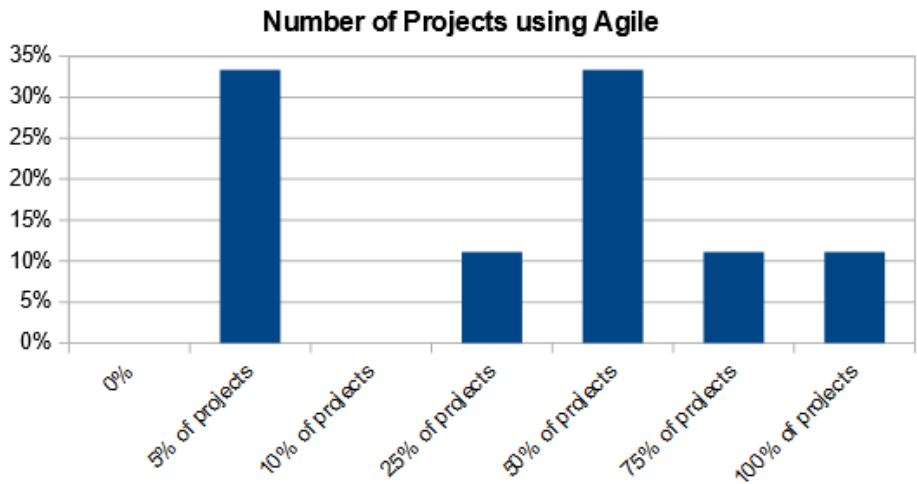

**Fig. 16.** Number of Projects using Agile (Own authorship)

## 5    Discussion

When analyzing the results obtained and present in the figures in the previous section, we can see that it was possible to identify similarity with the study conducted by VersionOne, mainly in the difficulties for the adoption of agile methods: organizational culture in disagreement with agile values (63%) and lack of skills or experience with agile methods (47%). This study shows that the main barrier to the full adoption of agile methods is the lack of experience (66.66%). One of the possible causes may be the lack of training in agile methods and techniques, according to the opinions expressed by people related to the agile community in Paraguay.

In three aspects our results were very similar to those obtained in [13]: total size of the technology team, time of experience of the company in agile methods and most used method (Scrum). The main differences were related to:

- Percentage of projects carried out with agile methods. In [13], 30.4% of the companies developed all of their projects using agile methods. In our study, 11.11% of companies do the same;

- Profile of participants in the survey. In [13] 18.5% of respondents were developers. In our study, by the very design of the research, no developer replied.

The greatest concern for the initial adoption reflects the following: 44.44% of participants had concerns about software quality at the time they adopted the new method. The other reasons are: inability to climb, with 22.22% and development team resistant to changes, with 22.22%. With respect to software quality, it can be deduced as a lack of knowledge or further training under the methods and techniques, because what is nailed is precisely a better alignment with the client so that what is delivered is closer than expected for the client.

It is important to note that the results can not be generalized statistically because it corresponds to a preliminary study that aims to be complemented with larger data that may be significant and allow a concrete visualization of the mentioned scenario.

## 6    Conclusions

This research was carried out with the purpose of identifying the level of adoption of agile methods in software development companies in Paraguay, raising the barriers and the concerns for their implementation. The answers to the questionnaire reveal that these companies experience the use of methods and techniques, and the main concerns they reported are (a) reduced software quality, (b) change resistant development team, and (c) inability to scale.

The barriers reported are (a) little experience, (b) confidence in the ability to scale agile methods, (c) little or no customer collaboration, and (d) complexity or size of projects. Another interesting result is that more than 50% of the companies adopt the Scrum Framework.